%% file: main.tex
\documentclass[runningheads]{llncs}
\usepackage[utf8]{inputenc}
\usepackage{graphicx}
\usepackage{amsmath}
\usepackage{floatrow}
\usepackage{booktabs}
\usepackage{tabularx}
\usepackage{hyperref}
\usepackage{caption}
\usepackage{subcaption}
\usepackage{xcolor}
\newcommand{\shrink}{\vspace*{-.9\baselineskip}}
\newcommand{\sectionshrink}{\vspace*{-.75\baselineskip}}

\begin{document}
\title{Graph-Embedding Empowered Entity Retrieval}
%
%
%
	\author{Emma J. Gerritse
\and Faegheh Hasibi \and
Arjen P. de Vries}
\authorrunning{E.J. Gerritse et al. }
%
\institute{  Institute for Computing and Information Sciences\\ Radboud University, Nijmegen, the Netherlands \\
\email{emma.gerritse@ru.nl f.hasibi@cs.ru.nl a.devries@cs.ru.nl} }
\maketitle              
\begin{abstract}
In this research, we improve upon the current state of the art in
entity retrieval by re-ranking the result list using graph
embeddings. The paper shows that graph embeddings are useful for
entity-oriented search tasks. We demonstrate empirically that encoding
information from the knowledge graph into (graph) embeddings
contributes to a higher increase in effectiveness of entity retrieval
results than using plain word embeddings. We analyze the impact of the
accuracy of the entity linker on the overall retrieval
effectiveness. Our analysis further deploys the cluster hypothesis to
explain the observed advantages of graph embeddings over the more
widely used word embeddings, for user tasks involving ranking
entities.
\end{abstract}
\input{introduction}

\input{relatedwork}

\input{methods}
\input{experiment}

\input{results}

\input{conclusion}

\appendix
\input{appendices}

%

%
%
%
\bibliographystyle{splncs04}

\bibliography{literature}

\end{document}

%% file: introduction.tex
\section{Introduction}
\label{sec:intro}


Many information needs are entity-oriented, and with the rise of
knowledge graphs in Web and enterprise search~\cite{noy2019industry},
the role of entities has gained importance, both in the UI/UX where
so-called entity cards are shown in response to entity-oriented
queries, and in the ranking, where presence and absence of entity
mentions is weighted differently from traditional term occurrences.

Recently, word embeddings have been shown to be helpful for a number
of information retrieval problems. In the case of entity retrieval, a
natural representation would however not just represent words in
context of their textual neighborhood, but in context of the
knowledge graph instead. Here, we would want to apply graph embeddings
instead of word embeddings, where the semantic space constructed by
graph embeddings does not only encode the textual context of an entity
mention, but also the context as defined through the knowledge graph.
Considering Wikipedia as the knowledge graph to define the entities of
interest, for example, creating a graph embedding representation does
not just take the entity's page itself as context, but also its
anchor text, presence in lists and/or tables, \emph{etc.} 
It is therefore likely that graph embeddings capture more of the
entity's semantic roles and as a result may distinguish better between 
ambiguous entities than a plain word embedding based representation.

Exploring the use of graph embeddings in entity retrieval, we have
studied a two-stage entity retrieval approach where the second stage
employs graph embeddings for re-ranking the retrieval results of
state-of-the-art entity ranking methods. We investigate the following
research questions:

\begin{description}
\item[RQ1:] Does adding graph embeddings improve entity retrieval methods?
\item[RQ2:] Which queries are helped the most?
\end{description}

To our knowledge, we are the first to investigate how the structural
information captured in graph embeddings can contribute to improved
retrieval effectiveness in entity-oriented search.
The contributions of this paper are as follows:
We have build graph embeddings from Wikipedia as a knowledge graph%
\footnote{Downloadable at \url{https://github.com/informagi/GEEER}}
and evaluated the contribution of these embeddings as a representation
of entities in the ranking algorithm, using the DBpedia-Entity V2
collection~\cite{Hasibi:2017:DVT}. For every query, we re-rank the results of state-of-the-art
entity retrieval methods using the similarity between the entity
embeddings of the candidate entities retrieved in stage one with the
entity embeddings of the entities identified in the query (using an
off-the-shelf entity linker). We show that re-ranking using graph
embeddings improves retrieval effectiveness, and investigate how to
explain this result by comparing the structure of the two types of
embeddings.
We also analyze why some queries are helped by this method while
others are not.



%% file: relatedwork.tex
\section{Related Work}
\label{sec:related}

\subsection{Word and Graph Embeddings}

Distributional representations of language have been object of study
for many years in natural language processing (NLP), because of their promise to represent words not in isolation, but `semantically', with their immediate context. 
Algorithms like Word2Vec~\cite{Mikolov:2013:DRW} and Glove~\cite{pennington2014glove} construct a vector space of word domains where similar words are mapped together (based on their linguistic context).
Word2Vec uses neural networks to predict words based on the context (continuous bag of words) or context based on a word (skip gram).
These word embedding representations have turned out to be highly effective in a wide variety of NLP tasks.

Word embeddings have been shown to help effectiveness in document retrieval\ \cite{dehghani2017neural,diaz2016expansion}. 
In~\cite{diaz2016expansion}, locally trained word embeddings are used for query expansion. 
Here queries are expanded with terms highly similar to the query, and it is shown that this method beats several other neural methods. In~\cite{dehghani2017neural}, embeddings are used for weak supervision of documents. 
This paper uses query embeddings and document embeddings to predict relevance between queries and documents, when given BM25 scores as labels. It is able to improve on BM25.  

Word embeddings consider the immediate linguistic context of the word occurrences. 
Going beyond just the text itself, researchers have proposed to develop so-called \emph{graph embeddings} to encode not just words in text, but words in context of semi-structured documents
represented as graphs - for example, to distinguish the occurrence of a word in the title of a document from its occurrences in a paragraph, or in a document's anchor text.

Different methods to produce graph embeddings have been proposed.
Methods like Deepwalk~\cite{perozzi2014deepwalk} expect non-labeled edges and can be considered extensions of the word embedding approaches discussed before. 
Other approaches include the well-known method Trans-E~\cite{bordes2013translating}, where edges in the graph are denoted as triples \textit{(head, label, tail)}, where \textit{label} is the
value of the edge. 
Adding graph embedding vectors of the \textit{head} and the \textit{label} should result in the vector of the \textit{tail}. 
The embeddings here are learned by gradient descent.

Wikipedia2Vec~\cite{Yamada:2016:JLE} applies graph embeddings to Wikipedia, creating embeddings that jointly capture link structure and text. 
The Wikipedia knowledge graph is indeed a natural resource for using graph embeddings, because it represents entities in a graph of interlinked Wikipedia pages and their text. The method proposed in~\cite{Yamada:2016:JLE} embeds words and entities in the same vector space by using word context and graph context. 
The word-word context is modeled using the Word2Vec approach, entity-entity context considers neighboring entities in the link graph, and word-entity context takes the words in the context of the anchor that links to an entity. 
The authors of Wikipedia2Vec demonstrate performance improvements on a
variety of NLP tasks, although they did not consider entity retrieval in their work.

\subsection{Entity retrieval}\label{sec:er}

An entity is an object or concept in the real world that can be distinctly identified~\cite{balog2018entity}. Knowledge graphs
like Wikipedia enrich the representation of entities by modeling the
relations between them. Methods for document retrieval such as BM25
have been applied successfully to entity retrieval. However, since
knowledge bases are semi-structured resources, this structural
information may be used as well, for example by viewing entities as
fielded documents extracted from the knowledge graph. A well-known
example of this approach applies the fielded probabilistic model
(BM25F~\cite{robertson2009probabilistic}), where term frequencies
between different fields in documents are normalized to the length of
each field. Another effective model for entity retrieval uses the
fielded sequential dependence model (FSDM~\cite{zhiltsov2015fielded}),
which estimates the probability of relevance using information from
single terms and bigrams, normalized per field.

\subsection{Using entity linking for entity retrieval}\label{sec:elr}

Linking entities mentioned in the query to the knowledge graph~\cite{Blanco:2015:FSE,Hasibi:2015:ELQ} enables
the use of relationships encoded in the knowledge graph, helping
improve the estimation of relevance of candidate entities. Previous
work has shown empirically that entity linking can increase
effectiveness of entity retrieval. In~\cite{hasibi2016exploiting}, for
example, entity retrieval has been combined with entity linking to
improve retrieval effectiveness over state-of-the-art methods
like FSDM.

Our research uses the \textsc{Tagme} entity linker~\cite{ferragina2010tagme} because it is especially suited to annotate
short and poorly composed text like the queries we need to link
to. \textsc{Tagme} adds Wikipedia hyperlinks to parts of the text,
together with a confidence score.

\subsection{Using embeddings for entity retrieval}

Very recent work has applied Trans-E graph embeddings to the problem
of entity retrieval, and shown consistent but small improvements~\cite{liu2019explore}. However, Trans-E graph embeddings are not a
good choice if the graph has 1-to-many, transitive or symmetric
relations, which is the case in knowledge graphs~\cite{paulheimpresentation}. In our research, we also look into
improving entity retrieval using graph embeddings, but use the
Wikipedia2Vec representation to address these shortcomings.

%% file: methods.tex
\section{Embedding Based Entity Retrieval}
\label{sec:eer}


\subsection{Graph Embeddings}
\label{sec:eer:embeddings}

We base the training of our entity embeddings on  Wikipedia2Vec~\cite{Yamada:2016:JLE,Yamada:2018:WOT}. Taking a knowledge graph as the input, Wikipedia2Vec extends the skip-gram variant of Word2Vec~\cite{Mikolov:2013:DRW,Mikolov:2013:EEW} and learns word and entity embeddings jointly. The objective function of this model is composed of three components.
%
The first component infers optimal embeddings for words $W$ in the corpus. Given a sequence of words $w_1 w_2 ... w_T$ and a context window of size $c$, the word-based objective function is:
%
\begin{equation}
  \mathcal{L}_w = \sum_{t = 1}^T \sum_{-c \leq j \leq c, j \neq 0}
  \log \frac{\exp(\mathbf{V}_{w_t}^T \mathbf{U}_{w_{t+j}})}{\sum_{w\in W}
    \exp(\mathbf{V}_{w_t}^T\mathbf{U}_{w})}, 
\end{equation}  
%
where matrices $\textbf{U}$ and $\textbf{V}$ represent the input and output vector representations, deriving the final embeddings from matrix $\textbf{V}$.

The two other components of the objective function take the knowledge graph into account. One addition considers a link-based measure estimated from the knowledge graph (i.e., Wikipedia). This measure captures the relatedness between entities in the knowledge base, based on the similarity between their incoming links:
\begin{equation}
  \mathcal{L}_e = \sum_{e_i\in \mathcal{E}} \sum_{e_o \in C_{e_i}, e_i \neq e_i}
  \log \frac{\exp(\mathbf{V}_{e_i}^T\mathbf{U}_{e_o})}{\sum_{e\in
      \mathcal{E}}\exp(\mathbf{V}_{e_i}^T\mathbf{U}_{e})}.
\end{equation} 
Here, $C_e$ denotes entities linked to an entity $e$, and $\mathcal{E}$ represents all entities in the knowledge graph.

The last addition to the objective function places similar entities
and words near each other by considering the context of the anchor
text. The intuition is the same as in classic Word2Vec, but here,
words in the vicinity of the anchor text have to predict the
entity mention.
Considering a knowledge graph with anchors $A$ and an entity $e$ the goal is to predict context words of the entity:
\begin{equation}
  \mathcal{L}_a = \sum_{e_i \in A} \sum_{w_o \in a(e_i)}
  \log\frac{\exp(\mathbf{V}_{e_i}^T\mathbf{U}_{w_o})}{\sum_{w\in W}\exp(\mathbf{V}_{e_i}^T\mathbf{U}_{w})}, 
\end{equation} 
where $a(e)$ gives the previous and next $c$ words of the referent entity $e$.

These three components (word context, link structure, and anchor context)
are then combined linearly into the following objective function:
\begin{equation}
  \label{eq:obj}
  \mathcal{L} = \mathcal{L}_w + \mathcal{L}_e + \mathcal{L}_a.
\end{equation}

\subsection{Re-ranking Entities}
\label{sec:eer:er}

Training the Wikipedia2Vec model on a Wikipedia knowledge graph
results in a single graph embedding vector for every Wikipedia
entity. The next question to answer is how to use these
graph embeddings in the setting of entity retrieval. 

We propose a two-stage ranking model, where we first produce a ranking
of candidate entities using state-of-the-art entity retrieval models
(see Section \ref{sec:er}), and then use the graph embeddings to
reorder these entities based on their similarity to the query
entities, as measured in the derived graph embedding space. 

Following the related work discussed in Section \ref{sec:elr}, we use
the \textsc{Tagme} entity linker to identify the entities mentioned in
the query. Given input query $Q$, we obtain a set of linked entities
$E(Q)$ and a confidence score $s(e)$ for each entity, which represents
the strength of the relationship between the query and the linked
entity. We then compute an embedding-based score for every query $Q$ and
entity $E$:
\begin{equation}\label{eq:score}
F(E,Q) = \sum_{e\in E(Q)} s(e) \cdot cos(\overrightarrow{E}, \overrightarrow{e}),
\end{equation}
where $\overrightarrow{E}, \overrightarrow{e}$ denote the embeddings
vectors for entities $E$ and $e$.

The rationale for this approach is the hypothesis that relevant
entities for a given query are situated close (in graph embedding
space) to the query entities identified by the entity linker.

Consider for example the query \textit{``Who is the daughter of Bill
  Clinton married to.''} \textsc{Tagme} links the query to entities
\textsc{Bill Clinton} with a confidence of $0.66$, \textsc{Daughter}
with a confidence of $0.13$, and \textsc{Same-sex marriage} with a
confidence score of $0.21$. Highly ranked entities then have a large
similarity to these entities, where similarity to \textsc{Bill
  Clinton} adds more to the score than similarity to \textsc{Daughter}
or \textsc{Same-sex marriage} (as the confidence score of \textsc{Bill
  Clinton} is higher than the other two).
The relevant entities for this query (according to the DBpedia-Entity
V2 test collection~\cite{Hasibi:2017:DVT}) are \textsc{Chelsea
  Clinton}, who is Bill Clinton's daughter, and \textsc{Clinton
  Family}. We can reasonably expect these entities to have similarity
to the linked entities, confirming our intuition. 

To produce our final score, we interpolate the embedding-based score
computed using Eq.~\eqref{eq:score} with the score of the
state-of-the-art entity retrieval model used to produce the candidate
entities in stage one: 
\begin{align}
	\label{eq:lincomb}
  &\mathit{score}_{\mathit{total}}(E,Q) = (1-\lambda) \cdot \mathit{score}_{\mathit{other}} (E, Q) + \lambda \cdot F(E,Q)
  & \lambda \in [0,1].
\end{align}

%% file: experiment.tex
\section{Experimental Setup}
\label{sec:exp}


\subsection{Test collection}
\label{sec:exp:coll}
In our experiments, we used the DBpedia-Entity V2 test
collection~\cite{Hasibi:2017:DVT}. The collection consists of 467
queries and relevance assessments for 49280 query-entity pairs, where
the entities are drawn from the DBpedia 2015-10 dump. The relevance
assessments are graded values of 2, 1, and 0 for highly relevant,
relevant, and not relevant entities, respectively. The queries are
categorized into 4 different groups: \textbf{SemSearch ES}  consisting
of short and ambiguous keyword queries (e.g.,\emph{``Nokia E73''}),
\textbf{INEX-LD} containing IR-Style keyword queries (e.g.,
\emph{``guitar chord minor''}), \textbf{ListSearch} consisting of
queries seeking for a list of entities (e.g., \emph{``States that
  border Oklahoma''}), and \textbf{QALD-2} containing entity-bearing
natural language queries (e.g., \emph{``Which country does the creator
  of Miffy come from''}). Following the baseline runs curated with the
DBpedia-Entity V2 collection, we used the stopped version of queries,
where stop patterns like ``which'' and ``who'' are removed from the
queries.

%
%
\subsection{Embedding Training}
\label{sec:exp:training}

Wikipedia2Vec provides pre-trained embeddings. 
These embeddings, however, are not available for all
entities in Wikipedia; e.g., 25\% of the assessed entities in
DBpedia-Entity V2 collection have no pre-trained embedding. The
reasons for these missing embeddings are two-fold: (i) ``rare''
entities were excluded from the training data, and, (ii) entity
identifiers evolve over time, resulting in entity mismatches
with those in the DBpedia-Entity collection.

For training new graph embeddings, we used Wikipedia 2019-07 dump. This was the newest version at the time of training. We
address the entity mismatch problem by identifying the entities that
have been renamed in the new Wikipedia dump. Some of these entities
were obtained using the redirect API of
Wikipedia.\footnote{\url{https://wikipedia.readthedocs.io/en/latest/}}
Others were found by matching the Wikipedia page IDs of the two
Wikipedia dumps. The page IDs of Wikipedia 2019-07 were available on
the Wikipedia website. For the dump where DBpedia-Entity is based on, however, these IDs are
not available anymore; we obtained them from the Nordlys
package~\cite{Hasibi:2017:NTE}.

To avoid excluding rare entities and generate embeddings for a wide
range of entities, we changed several Wikipedia2Vec settings. The
two settings that resulted in the highest coverage of entities are:
(i) minimum number of times an entity appears as a link in Wikipedia,
(ii) whether to include or exclude disambiguation
pages. Table~\ref{tbl:missingent} shows the effect of these settings
on the number of missing entities; specifically the number of entities
that are assessed in the DBpedia-Entity collection, but have missing
embeddings. We categorize these missing entities into two groups:
 %
\begin{itemize}
	\item \emph{No-page}: Entities without any pages. These entities neither were found by the Wikipedia redirect API nor could be matched by their page IDs. 
	\item \emph{No-emb}: Entities that could be found by their identifiers, but were not included in the Wikipedia2Vec embeddings.
\end{itemize}

The first line in Table~\ref{tbl:missingent} corresponds to the
default setting of Wikipedia2Vec, which covers only 75\% of assessed
entities in the DBpedia-Entity collection. When considering all entities
in the knowledge graph, this setting discards an even larger number of
entities, which is not an ideal setup for entity ranking. By choosing
the right settings (the last line of Table~\ref{tbl:missingent}), we
increased the coverage of entities to 97.6\%.

We trained two versions of embeddings: with and without link graph;
i.e., using Eq.~\eqref{eq:obj} with and without the $\mathcal{L}_e$
component.


%

\setlength{\arrayrulewidth}{1.5\arrayrulewidth}
\begin{table}[t]
	\centering
	\caption{Missing entities with different settings}
	\begin{tabularx}{0.9\linewidth}{| X | r | r | r |}
		\hline
		\textbf{Settings} & \textbf{No-emb} & \textbf{No-page} & \textbf{Total} \\
		\hline
		min-entity-count = 5, disambiguation = False & 9640    & 608    & 10248 \\
		min-entity-count = 1, disambiguation = False & 1220    & 398    & 1618  \\
		min-entity-count = 1, disambiguation = True & 1220    & 377    & 1597  \\
		min-entity-count = 0,  disambiguation = False                  & \textbf{724} & 380 & 1104  \\
		min-entity-count = 0,  disambiguation = True  & \textbf{724} & \textbf{333} & \textbf{1057} \\
		\hline
	\end{tabularx}
	\label{tbl:missingent}
\end{table}

\subsection{Parameter Setting}
\label{sec:exp:param}
Our entity re-ranking approach involves free parameter $\lambda$ that
needs to be estimated (see Eq.~\eqref{eq:lincomb}). To set this 
parameter, we employed the Coordinate Ascent
algorithm~\cite{Metzler:2007:LFM} with random restart of 3,
optimized for NDCG@100. All experiments were performed using
5-fold cross-validation, where the folds were obtained from the
collection (DBpedia-Entity V2). This makes our results comparable to
the DBpedia-Entity V2 baseline runs, as the same folds are used for
all the methods. Entity re-ranking was performed on top 1000
entities ranked by two state-of-the-art term-based entity retrieval
models: FSDM and BM25F-CA~\cite{Hasibi:2017:DVT}. For all experiments,
we used the embedding vectors of 100 dimensions, which were trained
using the settings described in Section~\ref{sec:exp:training}.

%% file: results.tex
\section{Results and Analysis}
\label{sec:results}

\setlength{\arrayrulewidth}{1.5\arrayrulewidth}
\begin{table*}
  \centering
  \footnotesize
  \caption{Results of embedding-based entity re-ranking approach on different query subsets of DBpedia-Entity V2 collection.  Significance of results is explained in running text.} 
  \label{tbl:res}
  \begin{tabular}{|l @{~}|| @{~}r@{~~}r | r@{~~}r | r@{~~}r | r@{~~}r | r@{~~}r|}
    \hline
    {\textbf{Model}} & \multicolumn{2}{c|}{\textbf{SemSearch}} & \multicolumn{2}{c|}{\textbf{INEX-LD}} & \multicolumn{2}{c|}{\textbf{ListSearch}} & \multicolumn{2}{c|}{\textbf{QALD-2}} & \multicolumn{2}{c|}{\bfseries Total}\\
     NDCG & @10 & @100 & @10 & @100 & @10 & @100 & @10 & @100 & @10 & @100 \\
    \hline
    \multicolumn{11}{|l|}{\emph{Reranking the FSDM top 1000 entities}}\\
    \hline
    $\text{ESim}_c$     & 0.365 & 0.412 & 0.194 & 0.252 & 0.210 & 0.288 & 0.192 & 0.255 & 0.239 & 0.300 \\      
    $\text{ESim}_{cg}$ & 0.397 & 0.462 & 0.216 & 0.282 & 0.211 & 0.311 & 0.213 & 0.286 & 0.258 & 0.334 \\   
   FSDM         & 0.652 & 0.722 & 0.421 & 0.504 & 0.420 & 0.495 & 0.340 & 0.436 & 0.452 & 0.534 \\
    ~ +ELR 	& 0.656 &	0.726 &	0.435 &	0.513 &	0.422 &	0.496 &	0.347 &	0.446 &	0.459 &	0.541 \\
    ~ +ESim$_c$     & 0.659 & 0.725 & 0.433 & 0.513 & 0.432 & 0.509 & 0.353 & 0.447 & 0.463 & 0.543 \\
    ~ +ESim$_{cg}$ & \textbf{0.672} & 0.733 &  0.440 & 0.528 & 0.424 & 0.507  &  0.349 & 0.451 & 0.465 & 0.549 \\
    \hline
    \multicolumn{11}{|l|}{\emph{Reranking the BM25F-CA top 1000 entities}}\\
    \hline
    ESim$_{c}$     & 0.381 & 0.424 & 0.194 & 0.253 & 0.211 & 0.283 & 0.192 & 0.252 & 0.243 & 0.301 \\ 
    ESim$_{cg}$ & 0.417 & 0.478 & 0.217 & 0.286 & 0.211 & 0.302 & 0.212 & 0.282 & 0.262 & 0.335 \\
    BM25F-CA         & 0.628 & 0.720 & 0.439 & 0.530 & 0.425 & 0.511 & 0.369 & 0.461 & 0.461 & 0.551 \\
    ~ +ESim$_{c}$     & 0.658 & 0.730 & 0.462 & 0.545 & 0.448 & 0.529 & 0.380 & 0.469 & 0.481 & 0.563 \\
    ~ +ESim$_{cg}$ & 0.660 & \textbf{0.736} & \textbf{0.466} & \textbf{0.552} & \textbf{0.452} & \textbf{0.535} & \textbf{0.390} & \textbf{0.483} & \textbf{0.487} & \textbf{0.572}\\
    \hline
  \end{tabular}
\end{table*} 
\setlength{\arrayrulewidth}{0.75\arrayrulewidth}

\subsection{Overall Performance}\label{sec:results:overall}

To answer our first research question, whether embeddings improve the
score of entity retrieval, we compare our entity re-ranking approach
with a number of baseline entity retrieval models. Table~\ref{tbl:res}
shows the results for different models with respect to NDCG@10 and
NDCG@100, the default evaluation measures for DBpedia-entity V2. In
this table, the embedding-based similarity component
(Eq.~\eqref{eq:score}) is denoted by \emph{ESim}, where \emph{c} and
\emph{cg} subscripts refer to the two versions of our entity
embeddings: without and with link graph.

The results of our method are presented for components ESim$_{c}$ and
ESim$_{cg}$ by themselves (i.e., $\lambda =1$ in
Eq.\ \eqref{eq:lincomb}), and also in combination with FSDM and
BM25F-CA. The mean and standard deviation of $\lambda$ found by the Coordinate Ascent algorithm over all folds are: $0.34 \pm 0.02$ for FSDM+ESim$_{c}$,
 $0.61 \pm 0.01$ for FSDM+ESim$_{cg}$, $0.81 \pm 0.03$ for BM25F-CA+ESim$_{c}$, and $0.88 \pm 0.00$ for BM25F-CA+ESim$_{cg}$.
The results show that the embedding-based scores alone do not perform
very well, however, when combining them with other scores, the
performance improves by a large margin. 
We determine the statistical significance of the difference in
effectiveness for both the NDCG@10 and the NDCG\-@\-100 values, using
the two-tailed paired t-test with $\alpha < 0.05$. The results show
that both versions of FSDM+ESim and BM25-CA+ESim models yield significant
improvements over FSDM and BM25-CA models (with respect to all
metrics), respectively. Also, FSDM\-+ESim$_{cg}$ improves significantly
over FSDM\-+ELR with respect to NDCG@100, showing that our embedding based
method captures entity similarities better than the strong entity ID
matching approach used in the ELR method.

When considering the query subsets, we observe that FSDM\-+ESim$_{cg}$
significantly outperforms FSDM for SemSearch and QALD queries with
respect to NDCG@10, and for INEX-LD queries with respect to
NDCG@100. Improvements over BM25F-CA were more substantial:
BM25F-CA+ESim$_{cg}$ brings significant improvements for all
categories (with respect to all metrics) except for SemSearch queries
for NDCG@100.


\subsection{Entity Embeddings Analysis}
\label{sec:results:embd}

\begin{figure*}[t]
	\centering
	\begin{subfigure}[t]{0.49\textwidth}
		\includegraphics[width=\textwidth]{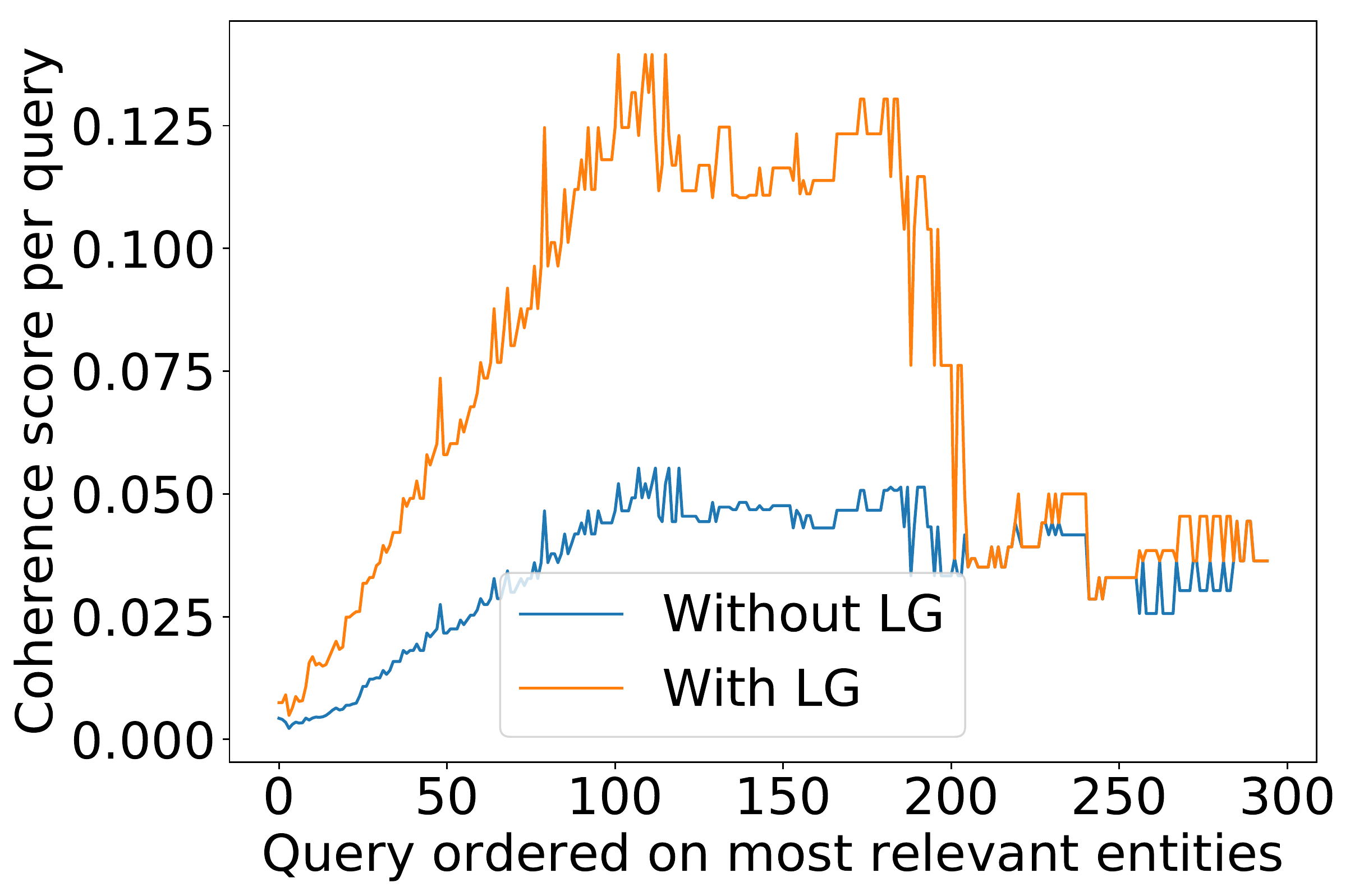}
		\caption{threshold $\tau=0.9$}
		\label{fig:coh:9}
	\end{subfigure}
	\begin{subfigure}[t]{0.49\textwidth}
		\includegraphics[width=\textwidth]{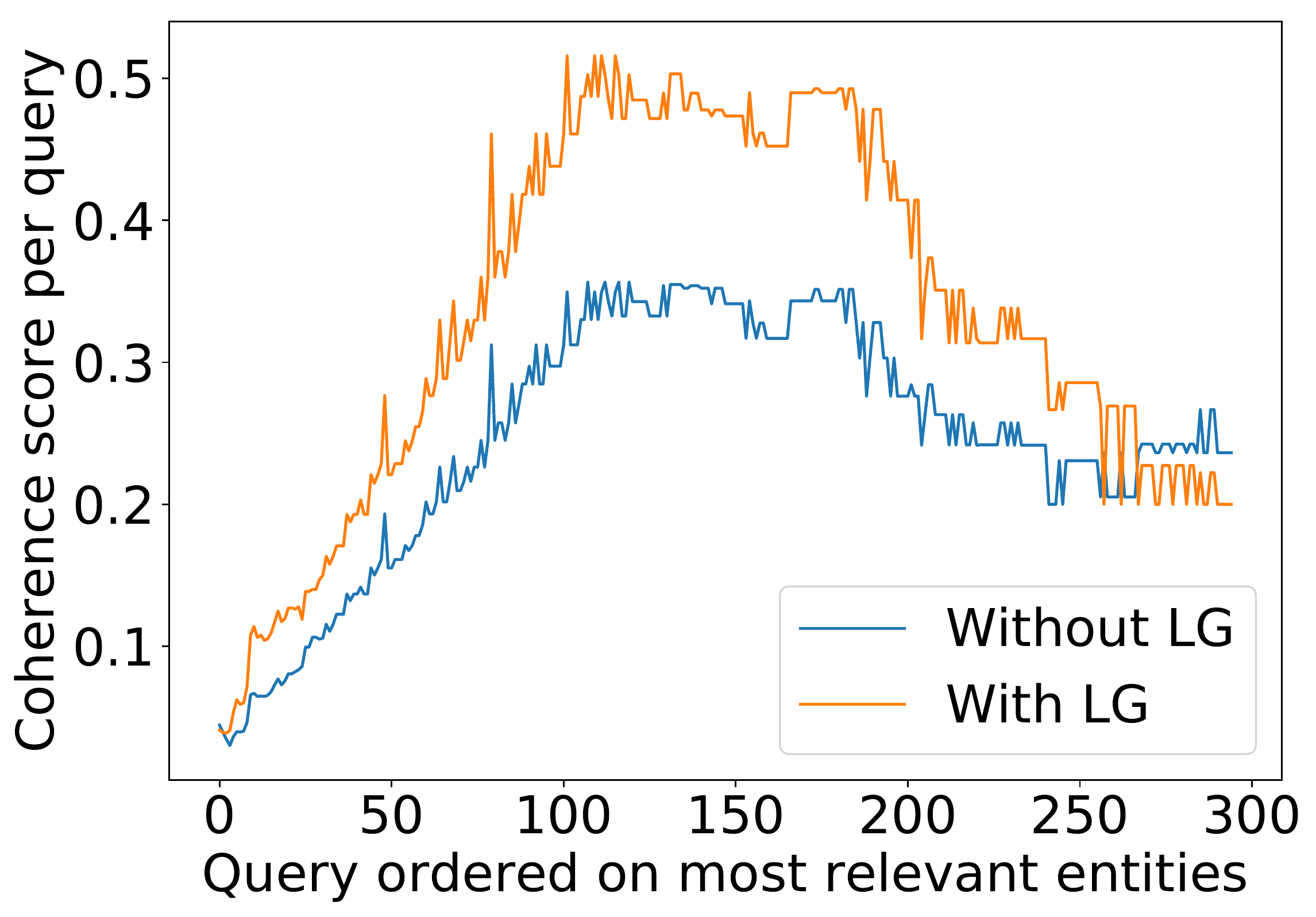}
		\caption{threshold $\tau=0.8$}
		\label{fig:coh:8}
	\end{subfigure}
	\caption{Coherence score of all relevant entities per query, computed for the versions of entity embeddings (without and with link graph) . The queries are ordered by the number of their relevant entities in x-axis.}
	\label{fig:coh}
\end{figure*}

The results of Table\ \ref{tbl:res} suggest that graph-based entity
embeddings yield better performance compared to context only entity
embeddings. To analyze why graph-based entity embeddings are
beneficial for entity retrieval models, we conduct a set of
experiments and investigate properties of embeddings with and without
the graph structure.


According to the cluster hypothesis~\cite{jardine1971use}, documents
relevant to the same query should cluster together. We consider
the embeddings as data-points to be clustered and compare the
resulting clusters in several ways. First, we compute the Davies
Bouldin index~\cite{davies1979cluster} and the Silhouette index~\cite{rousseeuw1987silhouettes}, which are: $3.16$ and $0.08$ for the
embeddings with link graph, and $3.98$ and $-0.05$ for the
embeddings without link graph, respectively. Both measures
indicate that better clusters arise for the embeddings that
capture graph structure.

To get an indication of how coherent the clusters are, we compute for
each query the coherence score defined in~\cite{he2011exploring}.
This score measures the similarity between item pairs of a cluster and
returns the percentage of items with similarity score higher than a
threshold, thereby assigning high scores to the clusters that
are coherent. Formally, given a document set $D$, the coherence score
is computed as:
\begin{equation}
	 Co(D) = \frac{\sum_{i\neq j \in {1, \dots, M} } \delta(d_i, d_j) }{\frac{1}{2} M(M-1)},
\end{equation}
where $M$ is total number of documents and the $\delta$ function for
each document pair $d_i$ and $d_j$ is defined as:
%
\begin{equation}
	\delta(d_i,d_j) =   
	\begin{cases}
		1, & \text{if}\ sim(d_i,d_j)\geq \tau \\
		0, & \text{otherwise}.
	\end{cases}
\end{equation}

%

We compute the coherence score with thresholds $0.8, 0.9$, using
$cosine$ for similarity function $sim(d_i, d_j)$, where $d_i$ and $d_j$
correspond to entities.
Figure~\ref{fig:coh} shows the results of coherence score for all queries
in our collection. Each point represents the coherence score of all
relevant entities (according to the qrels) for a query. We considered
only queries with more than 10 relevant entities, for clusters large
enough to compute a meaningful score. Queries are sorted on the x-axis
by the number of relevant entities.
The plots clearly show that the coherence score for graph-based
entity embeddings is higher than for context only ones. 
Based on these performance improvements we conclude that adding the
graph structure results in embeddings that are more suitable for
entity-oriented tasks.

\begin{figure*}[t]
	\centering
	\begin{subfigure}[t]{0.48\textwidth}
		\includegraphics[width=0.98\linewidth]{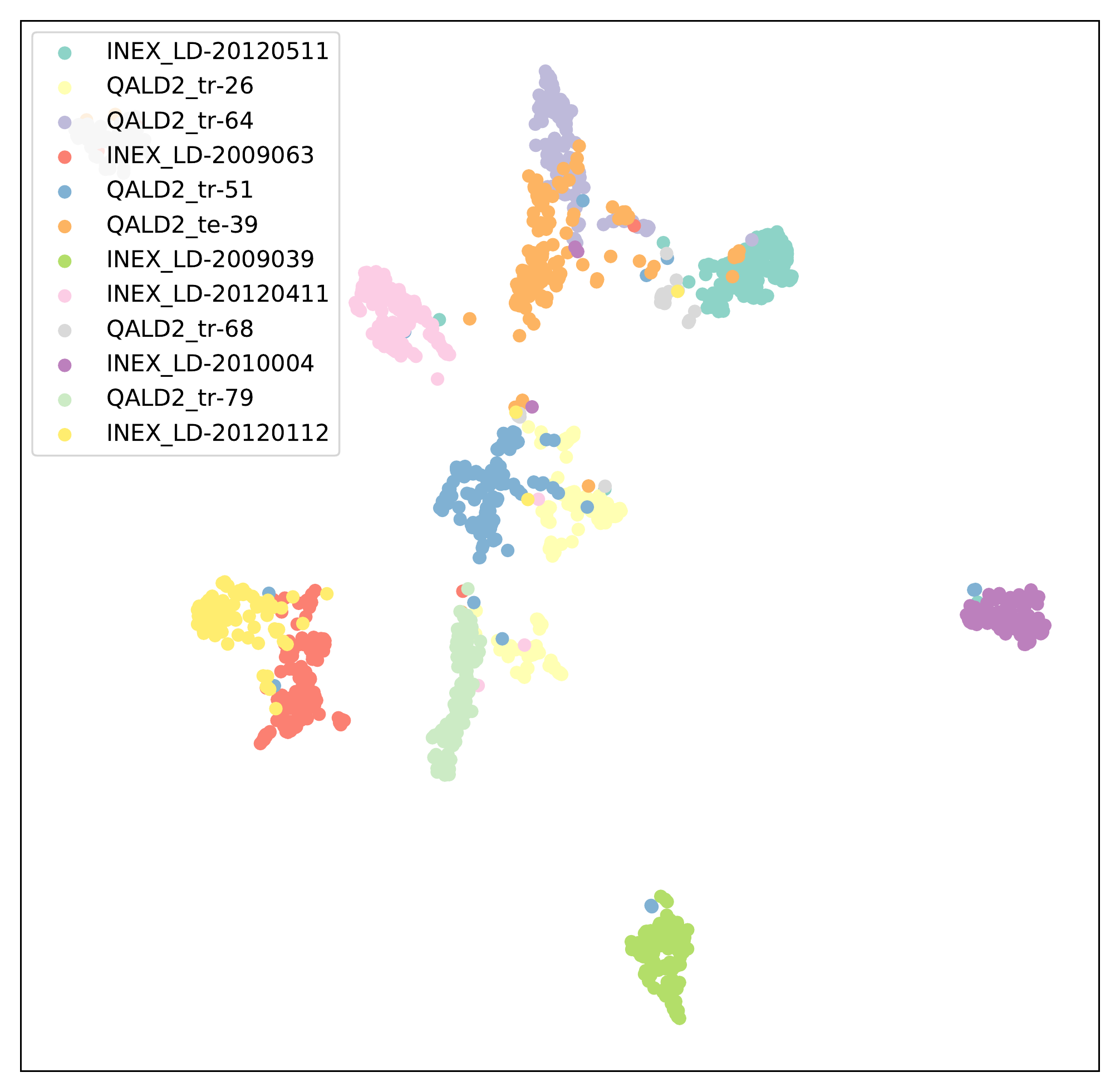}
		\caption{Embeddings with link graph}
		\label{fig:umap:lg}
	\end{subfigure}
	\begin{subfigure}[t]{0.48\textwidth}
		\includegraphics[width=\linewidth]{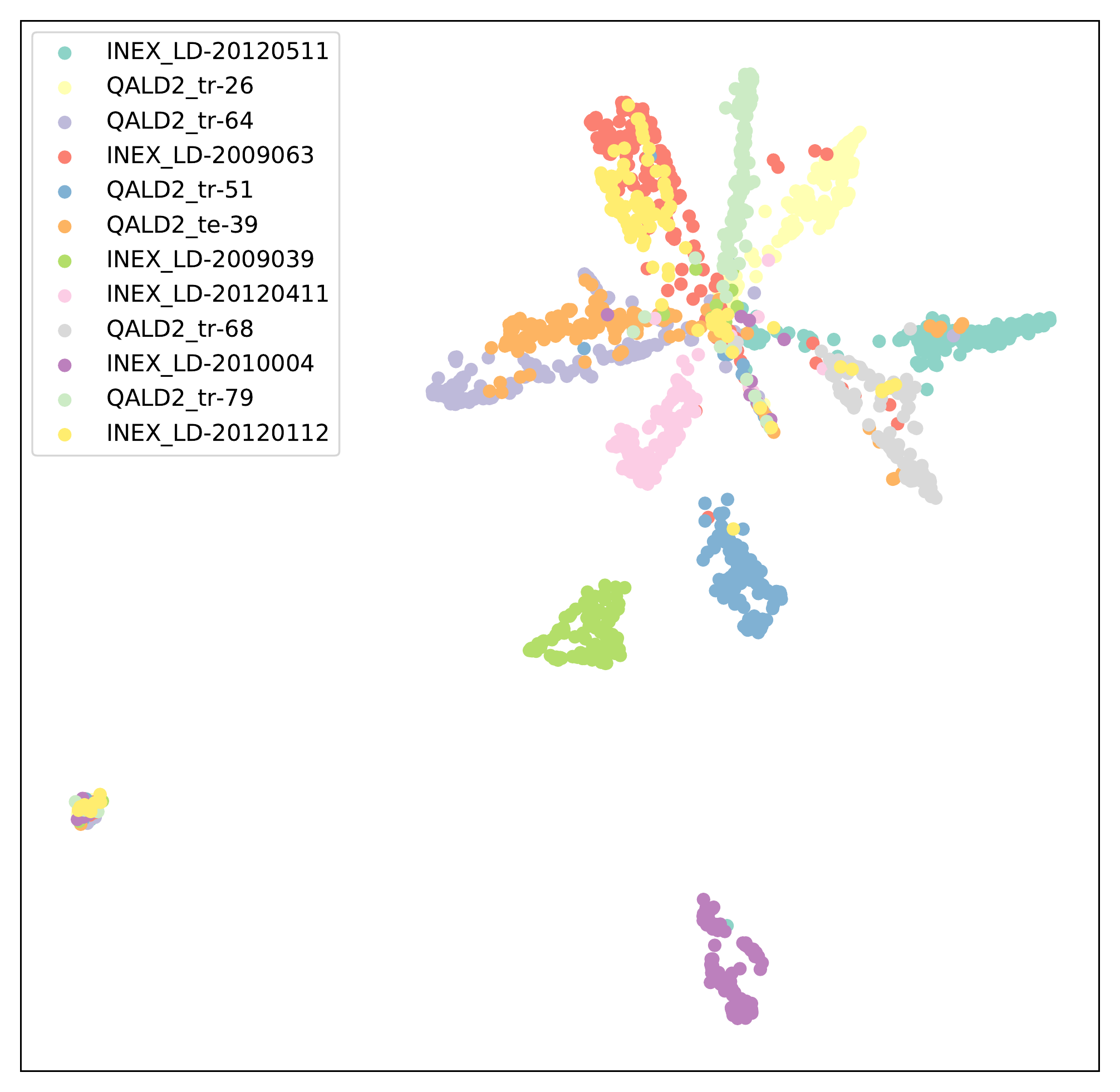}
		\caption{Embeddings without link graph}
		\label{fig:umap:nolg}
	\end{subfigure}
	\caption{UMAP visualization of entity embeddings for a subset of queries. Color-codes correspond to the relevant entities  per query. Queries per code are listed in Table~\ref{tbl:queries} of the Appendix. Default settings of UMAP in python were used.}
	\label{fig:umap}
\end{figure*}



Figures \ref{fig:umap} helps to visually understand how clusters of entities differ for the two methods (a subset of all entities is shown for clarity).
The data points correspond to the entities with a relevance grade higher than 0, for 12 queries
with 100--200 relevant entities in the ground truth data.
We use Uniform Manifold Approximation and Projection (UMAP)~\cite{mcinnes2018umap-software} to reduce the embeddings dimensions from 100 to two and plot the projected entities for each query.
In Figure~\ref{fig:umap:nolg} most of the clusters are overlapping in a star-like shape, while in Figure~\ref{fig:umap:lg} the clusters are more separated and the ones with similar search intents are close to each other; e.g., queries \texttt{QALD2\_te-39} and \texttt{QALD2\_tr-64} (which are both about companies), or \texttt{INEX\_LD-20120112} and \texttt{INEX\_LD-2009063} (which are both about war) are situated next to each other.
To observe how false positive entities are placed in the embedding space, we added the 10 highest ranked false positives to the data and created new UMAP plots. 
In the obtained plots, false positive entities that are semantically similar to the true positive entities are close to each other. For example, two false positive entities for the query \emph{``South Korean girl groups''} are: \textsc{SHINee} (a South Korean boy band) and \textsc{Hyuna} (a South Korean female singer). Both of these entities are semantically similar to the relevant entities of the query and are also placed in the vicinity of them, although they do not address the information needs of the query. This is consistent with the plots of Figure~\ref{fig:umap} and in line with our conclusion on the effect of graph embeddings for entity-oriented search.


\subsection{Query Analysis}
\label{sec:results:query}
\setlength{\arrayrulewidth}{1.5\arrayrulewidth}
\begin{table}[bt]
	\caption{Top queries with the highest gains and losses in NDCG at
		cut-offs 10 and 100, BM25F + ESim$_{cg}$ vs.\ BM25F.}
	\scriptsize
	\begin{tabularx}{0.9\linewidth}{|X|>{\raggedleft\arraybackslash}m{3em}|>{\raggedleft\arraybackslash}m{3em}|}
		\hline
		\multicolumn{1}{|c|}{Query} & \multicolumn{2}{c|}{Gain in NDCG}\\
		\hline
		& \multicolumn{1}{c|}{@10} & \multicolumn{1}{c|}{@100} \\
		\hline
		st paul saints                                                    & 0.716 & 0.482	  \\ 
		continents in the world                                           & 0.319 & 0.362	  \\
		What did Bruce Carver die from?                                   & 0.307 & 0.307	   \\
		\hline
		spring shoes canada                                               & -0.286 & -0.286 \\
		vietnam war movie                                                 & -0.470 & -0.240 \\
		mr rourke fantasy island                                          & -0.300 & -0.307 \\
		\hline
	\end{tabularx}
	\label{tab:bm25}
\end{table}
\setlength{\arrayrulewidth}{0.75\arrayrulewidth}
\setlength{\arrayrulewidth}{1.5\arrayrulewidth}
\begin{table}[bt]
	\caption{Top queries with the highest gains and losses in NDCG at
		cut-offs 10 and 100, BM25F + ESim$_{cg}$ vs.\ BM25F + ESim$_{c}$.}
	\scriptsize
	\begin{tabularx}{0.9\linewidth}{|X|>{\raggedleft\arraybackslash}m{3em}|>{\raggedleft\arraybackslash}m{3em}|}
		\hline
		\multicolumn{1}{|c|}{Query} & \multicolumn{2}{c|}{Gain in NDCG}\\
		\hline
		& \multicolumn{1}{c|}{@10} & \multicolumn{1}{c|}{@100} \\
		\hline
		What did Bruce Carver die from?                                   & 0.307 & 0.307	   \\
		Which other weapons did the designer of the Uzi develop?          & 0.236 & 0.248	  \\ 
		Which instruments did John Lennon play?                           & 0.154 & 0.200	  \\
		
		\hline
		Companies that John Hennessey serves on the board of              & -0.173 & -0.173 \\
		Which European countries have a constitutional monarchy?          & -0.101 & -0.197 \\
		vietnam war movie                                                 & -0.276 & -0.222 \\
		
		\hline
	\end{tabularx}
	\label{tab:bm25c}
\end{table}
\setlength{\arrayrulewidth}{0.75\arrayrulewidth}
%
Next, we investigate our second research question and analyse queries that are helped and hurt the most by our embedding-based method.
Table~\ref{tab:bm25} shows six queries that are affected the most by
BM25F-CA+ESim$_{cg}$ compared to BM25F-CA (on NDCG@100).
Each of the three queries with highest gains are linked to at least one
relevant entity (according to the assessments).
The losses can be attributed to various sources of errors.
For the query \emph{``spring shoe canada''}, the only relevant entity belongs to the 2.4\% of entities that have no embedding (cf.~\S\ref{sec:exp:training}). 
Query \emph{``vietnam war movie''} is linked to entities
\textsc{Vietnam War} and \textsc{War film}, with confidence scores of
0.7 and 0.2, respectively. This emphasizes Vietnam war facts instead
of its movies, and could be resolved by improving the
accuracy of the entity linker and/or employing a re-ranking approach
that is more robust to linking errors.
The query \emph{``mr rourke fantasy island''} is linked to a wrong
entity due to a spelling mistake. To conclude, errors in entity
linking form one of the main reasons of performance loss in our
approach.

To further understand the difference between the two versions of the embeddings at the query-level, we selected the queries with the highest and lowest gain in NDCG@100 (i.e., comparing BM25F+ESim$_{cg}$ and  BM25F+ESim$_c$).
For the query \emph{``Which instruments did John Lennon play?''}, the two linked entities (with the highest confidence score) are \textsc{John Lennon} and \textsc{Musical Instruments}. 
Their closest entity in graph embedding space is \textsc{John Lennon's musical instruments}, relevant to the query. 
This entity, however, is not among the most similar entities when we consider the context-only case.

For the other queries in Table \ref{tab:bm25c}, the effect is similar but less large than in the BM25F and BM25F + ESim$_{cg}$ case, probably due to the lower value of $\lambda$.


%% file: conclusion.tex
\section{Conclusion}
\label{sec:concl}

We investigated the use of entity embeddings for entity retrieval. We
trained entity embeddings with Wikipedia2Vec, combined these with
state-of-the-art entity ranking models, and find empirically that
using graph embeddings leads to increased effectiveness of query
results on DBpedia-Entity V2.

The empirical findings can be interpreted as evidence for the cluster
hypothesis. Including a representation of the graph structure in the
entity embeddings leads to better clusters and higher effectiveness of
retrieval results. We further see that queries which get linked to
relevant entities or pages neighboring to relevant entities get helped
the most, while queries with wrongly linked entities are helped the least.

We conclude that enriching entity retrieval methods with entity
embeddings leads to improved effectivenss, but acknowledge the
following limitations of this study. Not all query categories lead to
improvements on NDCG. While the state-of-the-art in entity-linking has
made significant progress in recent years, we applied \textsc{Tagme}
to identify the entities in queries. As we observed that lower
performance of queries can often be attributed to erroneously linked
entities, we expect better results by replacing this component for a
state-of-the-art approach. Finally, we have only experimented using
the embeddings constructed by Wikipedia2Vec, and plan to continue our
experiments using alternative entity embedding methods like TransE.


%% file: appendices.tex
\section{Queries}\label{app:queries}
\sectionshrink

\shrink
\begin{table*}
  \caption{Queries mentioned by their query ID.}\label{tbl:queries}
  \scriptsize
  \begin{tabularx}{\linewidth}{l@{~~~~}l@{~}}
    \toprule
    Query ID & Query text \\
    \midrule
    INEX\_LD-20120511 & female rock singers \\
    QALD2\_tr-26 & Which bridges are of the same type as the Manhattan Bridge? \\
    QALD2\_tr-64 & Which software has been developed by organizations founded in California? \\
    INEX\_LD-2009063 & D-Day normandy invasion \\
    QALD2\_tr-51 & Give me all school types. \\
    QALD2\_te-39 & Give me all companies in Munich. \\
    INEX\_LD-2009039 & roman architecture \\
    INEX\_LD-20120411 & bicycle sport races \\
    QALD2\_tr-68 & Which actors were born in Germany? \\
    INEX\_LD-2010004 & Indian food \\
    QALD2\_tr-79 & Which airports are located in California, USA? \\
    INEX\_LD-20120112 & vietnam war facts \\
    \bottomrule
  \end{tabularx}
  
\end{table*}